# Simultaneous Achievement of Driver Assistance and Skill Development in Shared and Cooperative Controls

Takahiro Wada

Ritsumeikan University

1-1-1 Noji-higashi, Kusatsu, Shiga, 525-8577, Japan

twada@fc.ritsume.ac.jp, Tel: +81-77-561-2798

Abstract:

Advanced driver assistance systems have successfully reduced drivers' workloads and increased safety. On the other hand, the excessive use of such systems can impede the development of driving skills. However, there exist collaborative driver assistance systems, including shared and cooperative controls, which can promote effective collaboration between an assistance system and a human operator under appropriate system settings. Given an effective collaboration setup, we address the goal of simultaneously developing or maintaining driving skills while reducing workload. As there has been a paucity of research on such systems and their methodologies, we discuss a methodology applying shared and cooperative controls by considering related concepts in the skill training field. Reverse parking assisted by haptic shared control is presented as a means of increasing performance during assistance, while skill improvement following assistance is used to demonstrate the possibility of simultaneous achievement of driver assistance through the reduction of workload and skill improvement.

1. Introduction

Driver assistance systems (DASs) such as adaptive cruise control (ACC), lane keeping assist systems (LKASs), and advanced emergency brake systems (AEBSs) have been developed to reduce drivers' workloads and mitigate collisions. The literature has reported negative changes in drivers' behavior with the introduction of DASs (Wilde 1998; Hoedemaeker and Brookhuis 1998; Abe and Richardson 2004), and researchers have reported on methods for designing systems, such as AEBSs, to minimize such behavioral adaptation (Hiraoka et al. 2011; Itoh et al. 2011). It has also been noted that the maintenance and growth of driving skills can be negatively impacted by the use of sophisticated driver-assistance systems (Tada et al. 2016). Following the definitions in SAE J3016 (SAE 2016), at automation levels of 4 or 5, drivers do not require driving intervention, while at level 3 or lower, intervention may be required and the driver remains responsible for safe vehicle operation. As it stretches current credibility to envision a situation in which all vehicles are operated at automated driving levels of 4 or 5, for the time being, drivers will be required to maintain or increase their driving skills.

Some DASs are cooperatively involved with the driver in driving. There are many types of DAS with cooperation, including control and cooperative control, which are related as discussed in the work of Flemisch et al. (2016) and Wada et al. (2016). It is also expected that collaborative DASs will be used in improving driver skills (Hirokawa et al. 2014). We can also conceive the possibility of achieving simultaneous development of driving skills and reduction of workload through appropriate system design, and there have, in fact, already been studies achieving both workload reduction and skill improvement in the case of, e.g., reverse parking (Tada et al. 2016).

If the main purpose of human-machine cooperation is to reduce the human driver's workloads, the automation introduced by a system should not contravene this goal. Thus, it is necessary to develop a specific methodology for increasing driver skills and simultaneously reducing his/her workload. In the field of rehabilitation research, it is known that motor skill improvement is significantly increased when task difficulty is tuned according to the user's skill level (Wada and Takeuchi 2008). In another context, eco-driving skills have been significantly improved when the level setting of an assist system was tuned on the basis of the drivers' skill levels to maintain workload at an appropriate level (Wada et al. 2011). These results suggest that the simultaneous achievement of workload reduction and skill improvement in the context of driver assistance can be achieved by employing the task difficulty adjustment methodology based on the adjustable features in human-machine cooperation techniques. Considerable effort has been made toward enhancing human skills (Druckman 1994; Salvendy 2006; Wickens and Hollands 2000); however, little is known about a methodology to achieve skill development in the use of DASs while decreasing the workload.

The purpose of this paper is to discuss a methodology to simultaneously achieve skill development and reduce the workload through the use of shared and cooperative controls based on concepts in the skill-training field. In Section 2, shared and cooperative controls are described. Section 3 takes the concepts and ideas behind training methods and uses them to develop techniques for simultaneously reducing workload and improving driving skills through the adaptation of skill training concepts in the training research field to the features of a cooperative DAS. In Section 4, we discuss an example of such a system for providing reverse parking assistance.

2. Human machine cooperation

2.1 Shared and cooperative controls

Collaboration between humans and machines is variously described as human machine cooperation, cooperative control, shared control, etc. Flemisch et al. (2016) postulated a framework of human-machine cooperation covering strategic, tactical, and operational levels, which corresponds to navigation, guidance, and control in the work of Sheridan (1992). In shared control, a human operator and an automated system achieve a single operational task via a single operation input, such as an automobile steering wheel (Abbink et al. 2012; Nishimura et al. 2015). Shared control is understood to involve physical control, with connections through, for instance, the steering operation established via haptics, referred to as haptic shared control (HSC) (Abbink et al. 2012; Nishimura et al. 2015). In this paper, we define cooperative control as collaborative work between human and machine over a wider range than shared control, with a human and an automated system working together to achieve tasks involving more than one maneuver, and the automated system supports subtasks of given tasks from a strategic, tactical, and control perspective, not limited to only a single maneuver. Under this definition, the cooperative control includes HSC.

2.2 Cooperative status in haptic shared control

2.2.1 Haptic shared control

In HSC, the vehicle control task is achieved cooperatively by a human and a DAS via a single operational input such as the steering mechanism. Fig. 1 shows a block diagram of the overall human-machine system engaging HSC, where $y$ denotes the lateral position of the vehicle as effected via lateral control. Two agents—the human driver and the DAS—cooperatively operate one plant (the steering mechanism) to achieve the desired vehicle motion.

Pseudo-works exerted on the steering mechanism by the driver and the DAS, are respectively defined as follows:

$$w_c(t) := \frac{1}{\Delta T} \int_{t-\Delta T}^{t} \tau_c(s) \dot{y} ds \qquad (1)$$

$$w_{das}(t) := \frac{1}{\Delta T} \int_{t-\Delta T}^{t} \tau_{das}(s) \dot{y} ds \qquad (2)$$

where $\tau_c$ and $\tau_{das}$ are the torques exerted on the steering mechanism by the human limb and the DAS, respectively, and $y$ denotes the lateral displacement of the vehicle. The scalar $\Delta T$ is the time window used for the work calculation. These are used for the evaluation of the cooperative status of the human driver and DAS in the next subsection.

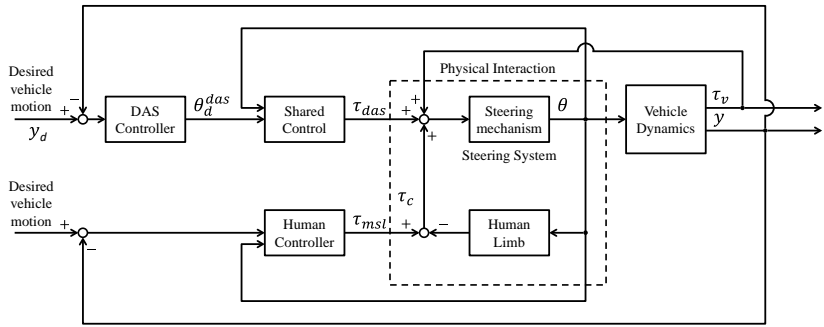

**Fig. 1** Block diagram of HSC. The human and DAS control the vehicle by sharing the steering operation. In effecting lateral control, $y$ denotes the lateral position of the vehicle on the road. The physical interaction between the human limb and steering mechanism is represented in the middle of the diagram.

The human controller determines the amount of muscle torque $\tau_{msl}$ to be applied from the feedback information provided by the vehicle motion and the current steering position $\theta$. The DAS controller outputs torque $\tau_{das}$ from the feedback information from the vehicle's motion. The scalar $\tau_c$ denotes the torque exerted on the steering wheel from the human limb. Refer to (Nishimura et al. 2015) for more detail on the dynamics.

2.2.2 Cooperative status

The human and the DAS—each having their own desired sets of motions—control a single vehicle using the controller. The cooperative status of both agents strongly affects the overall performance of controlling the vehicle as well as task difficulty experienced by the driver, as discussed in section 3.4.2. A methodology to evaluate the cooperative status between a human and an automated system using instant steering action and vehicle behaviour was developed from the viewpoints of a) initiative holder and b) intent consistency, which are defined as follows (Nishimura et al. 2015; Wada et al. 2016):

a) Initiative holder: The initiative holder is the agent currently having greater control of the vehicle motion. A human driver has the initiative when the following is satisfied:

$$w_c(t) \geq \gamma_1^2 \qquad (3)$$

where $\gamma_1^2$ is the offset of the judgment threshold.

b) Intent consistency: The intent consistency determines whether the human driver and DAS have the same operational intent. The intent of the two agents is consistent when the following is satisfied:

$$w_{das}(t) \geq \gamma_2^2 \quad \text{and} \quad w_c(t) \geq \gamma_1^2, \qquad (4)$$

where $\gamma_2^2$ is the offset of the judgment threshold. The intent of the two agents is defined to be inconsistent when the inequalities point in different directions.

Given the above two axes a) initiative holder and b) intent consistency, the cooperative status of the two agents is defined in Table 1 using $w_c(t)$ and $w_{das}(t)$.

State I: Driver-led cooperative state

In State I, the driver holds the initiative for vehicle operation in a cooperative manner with the assist control. This state occurs when both agents exert torque in the same direction and the vehicle moves in the intended direction.

State II: Driver-led uncooperative state

In State II, the driver holds the initiative for vehicle operation while the DAS attempts to steer against the driver. In this state, the vehicle moves in the driver's intended direction while the DAS exerts torque in the opposite direction.

State III: System-led state, which includes the following two sub-states:

III-a     System-led cooperative state

III-b     System-led uncooperative state

In III-a, the DAS holds the initiative for vehicle operation in a cooperative manner with the driver. In III-b, the DAS holds the initiative for vehicle operations while the driver attempts to steer against the DAS. It should be noted that it is difficult to distinguish between these two sub-states.

State IV: Passive state

This state rarely occurs over short intervals because of inertia or because a self-aligning torque is dominant.

State V: Dead zone

The blank area in Table 1 denotes a dead zone that is included to avoid misjudgments resulting from sensor noise.

A method to resolve intent inconsistency and achieve smooth transition of the cooperative status was presented by Nishimura et al. (2015).

**Table 1.** Cooperative states

| | | $w_c$ | | |
|---|---|---|---|---|
| | | $\leq -\gamma_1^2$ | else | $\geq \gamma_1^2$ |
| $W_{das}$ | $\geq \gamma_2^2$ | (III) System-led cooperative or uncooperative | | (I) Driver-led cooperative |
| | else | | (V) Dead zone | |
| | $\leq -\gamma_2^2$ | (IV) Passive: No active operation exists | | (II) Driver-led uncooperative: driver resisting DAS |

3. Simultaneous achievement of assistance for workload reduction and skill training

3.1 Training

Training is defined as "systematic acquisition of knowledge (what we need to know), skills (what we need to do), and attitude or ability (what we need to feel) (KSAs) that together lead to improved performance in a particular environment" (Salvendy 2006). Important features and methods to enhance training efficacy have been discussed as follows (Wickens et al. 2003; Wickens & Hollands 2000):

1) Practice and overlearning

The most basic and intuitive method for learning something involves practicing it repeatedly until performance reaches an expected goal (Anderson 1981; Fisk et al. 1987).

2) Reducing cognitive load

In some training processes, a large amount of instruction or other information is provided to the trainee.

Training is understood to be a process of transferring and storing acquired knowledge in any form into long-term memory and, according to cognitive load theory, it requires a certain amount of working memory (Sweller, 1994). In cases of overload, much information can be lost, resulting in low training efficacy. Thus, overload situations should be avoided (Sweller 1994).

3) Offering feedback/knowledge of results

Effective feedback is known to be essential for effective learning (Holding 1987). There are two types of feedback: corrective feedback, in which errors are noted; and motivational feedback, in which good performance of a given task is rewarded. Timely feedback is understood to be very important, with feedback ideally given immediately after a task is performed.

4) Encouraging active processing

Although it might be considered trivial, there is evidence that encouraging active participation is effective in learning (Goldman et al. 1999).

We focus on reducing cognitive load in the remaining part of paper. Several concepts are involved in reducing cognitive load:

a) Training in Parts

Training for complicated tasks may be difficult. In such cases, it can be useful to decompose the task into subtasks, practice each subtask in isolation, and then integrate them after they have been thoroughly trained for. This is called part-task training and can be classified into two types according to the method of dividing tasks as follows (Wightman and Lintern 1985). In *segmentation*, each sequential phase of a given task is practiced in isolation and then the learned parts are integrated into the whole. In *fractionization*, a given task is broken down into subtasks that are performed simultaneously but are practiced separately before combining into a whole. Segmentation is known to be useful in increasing training efficacy, while fractionization must be applied with caution because the advantage of the method is eliminated when the interdependence between subtasks is high. Overall, detailed task analysis is a key to success.

b) Simplifying

Simplifying is an approach to skill training in which training begins with simple tasks leading up to the task to be eventually performed. This would involve, for example, starting driving training at low speeds (Wightman and Lintern 1985). It is believed that simplifying can reduce both error, allowing the trainee to better engage in correct behavior, and cognitive load (Wickens et al. 2003). The method is thought to be achievable primarily in off-line training, in which the work environment can be controlled.

c) Guiding

Large errors can occur when a given task is difficult. For such cases, methods for giving assistance to reduce errors can be considered. Using the analogy of the auxiliary wheels of a child's bicycle, the approach is referred to as the training wheels approach (Carroll and Carrithers 1984).

d) Task difficulty adjustment (TDA)

This is a generic name for approaches in which the task difficulty is adjusted to a user's skill or ability when the given original task is too difficult to be accomplished by the user at first. Simplifying, guiding, and part-task training are all used in adjusting task difficulty. The method is known to enhance skill development in reverse parking (Wada et al. 2016) and eco-driving assistance systems (Wada et al. 2011) and, in the rehabilitation field, in the EMG prosthetic hand training method (Wada and Takeuchi 2008).

3.2 Effect of Task Difficulty Adjustment in Skill Training

TDA has following effects from the viewpoint of motor learning.

1) Feedback on success and failure: The user can experience a proper balance between success and failure, with feedback signals for positive and negative results enhancing motor learning ( Schmidt and Wrisberg 2007; Wickens and Hollands 2000).

2) Maintaining workload: The user's mental and physical workload is maintained at a certain level through the application of TDA adapted to the user's skill level. It is known that an appropriate workload has the effect of increasing task performance and active participation by the user in a given task, i.e., establishing a user-in-the-loop situation. In addition, as mentioned in the reduction of cognitive load approach, an appropriate setting of workload allows the trainee to use spare working memory to encode training results into their long-term memory or to enhance training efficacy.

3) Maintaining motivation: A user can attain self-efficacy by succeeding in tasks they consider challenging (Ryan and Deci 2000). Self-efficacy leads to maintained motivation for a given task. In addition, motivation is also known to be important in the transfer of training, meaning that skill improvement can be achieved even after the training or without the use of the assist systems used in the training phase (Wickens and Hollands 2000).

3.3 On the simultaneous achievement of driver support for workload reduction and skill development

As mentioned in Section 3.1, significant efforts have been made in developing effective training methods. However, there is a very limited understanding of which methodologies can be used to achieve both reduced workload during the training and skill development after deactivation of the assistance, simultaneously.

Here, we discuss a methodology to achieve workload reduction and skill development simultaneously in a driving context in which reducing the workload through TDA means reducing driving task difficulty via partial assistance with the task, which was the original aim of introducing the DAS. Here, we assume that the number of tasks is not changed with the assistance.

On the other hand, as mentioned in Section 2.1, the appropriate setting of task difficulty is important for skill development. Therefore, the two goals that we aim to achieve simultaneously can be described on a single axis as, for instance, the adjustment of task difficulty (Fig. 2) or workload. The key to accomplishing this is finding the task difficulty at which the workload is reduced and the skill is developed.

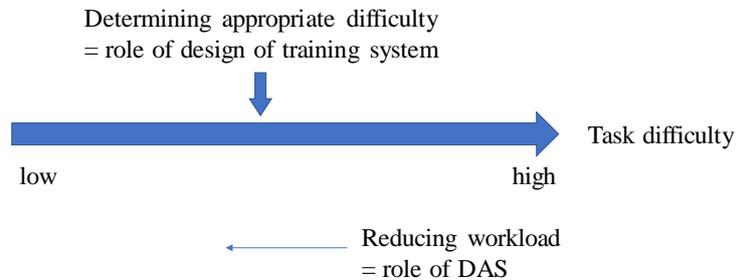

**Fig. 2** Driver assistance for workload reduction and skill training using the task-difficulty adjustment method can be achieved by appropriately adjusting task difficulty

The rehabilitation robotics field employs the error augmentation, or error amplification, approach in which the error to the given motor task is fed back via haptics with negative gain. Thus, the robot applies force in the direction in which the tracking error is enlarged (Wei et al. 2005; Patton et al. 2006). It is understood that this approach is effective in cases involving patients with relatively high skills. However, while failure or degrading of task performance with training is acceptable in rehabilitation because it is conducted separately from daily living situations, such Spartan training methods are not appropriate for driver assistance systems because such assistance systems are used in daily life and involve on-the-job training. A driver assistance system that increases driver workload is consequently considered unacceptable.

Developing a methodology of TDA suitable to the simultaneous achievement of workload reduction and skill development strongly depends on the methods employed in the driver assistance system. Consider a method for adjusting task difficulty in cooperative control. Effective cooperative control can be accomplished by knowing the operator's strengths and weaknesses, which in turn can be determined through analysis of the given task. When the extents of any weaknesses are identified, the assistance system can support the operator with the subtasks with which he or she requires assistance and allow him or her to perform subtasks that he or she is not good at but can still perform with an appropriate amount of effort. By maintaining the operator's workload appropriately through the use of this assistance strategy, the operator is expected to build up their skills in performing subtasks at which they are not good. Although no systematic method for adjusting task difficulty has been established, some ideas from a training viewpoint are given in the next subsection by type of DAS (Table 2).

3.4 Skill Training in cooperative control

The important factors in skill training—workload reduction, offering feedback, and encouraging active participation, as described in section 3.1—can be achieved by a collaborative DAS based on the nature of "cooperation" (F. Flemisch et al. 2016) in shared and cooperative controls.

Table 2 shows the relationship between the training concepts introduced in section 3.1 and the types of assistance in cooperative DAS and methods for adjusting task difficulty.

a) <u>Training in parts</u>

In the part-training approach, the human performs specified subtasks but not the whole task. In the context of driving, the DAS must perform the other subtasks required to achieve safe driving. There are several types of implementation methods for achieving this. In place of shared control, a technique called partitioning is discussed here.

The methods for partitioning of DAS are classified as follows:

(1) Fractionization into multiple tasks, meaning that the human and DAS take on different tasks entirely, as driving essentially comprises multiple subtasks involving lateral and longitudinal control (Fig. 3).

(2) Segmentation of tasks, meaning that different functions or stages of a task, such as decision making and control operation, are allocated to the human and DAS separately. This includes supporting decision-making and control (Fig. 4), which involves the DAS performing the decision-making and the driver enacting control operation based on the decision-making results, alternating with a mode in which the DAS performs control operation alone, based on the decision-making results of the human.

To adjust the task difficulty when using a partition-type DAS, tasks to be learned by the operator are set as the human's tasks when the task difficulty is appropriate for the individual. If the task difficulty is beyond the human's skill, additional assistance may be applied.

b) <u>Simplifying</u>

In the simplifying approach, the goal of the task is restricted or the training is started with a simplified goal according to the trainee's skill. There are two types of DAS with the simplifying approach:

(1) Giving support when difficult, meaning that the DAS gives support only if it judges that the given task is very difficult for the human operator. This should result in reduction in the difficulty in the driver's experience.

(2) Simplifying the traffic environment, meaning that the DAS attempts to control the traffic environment with the goal of decreasing its difficulty by operating or recommending future vehicle positions (Fig. 5). An example involving merging operation assistance is given in section 3.5.2.

After defining the difficulty in advance, it is adjusted by simply setting the threshold of the DAS, for example, when support should be given.

c) Guiding

In the guiding approach, shared control is primarily used to reduce errors for a given task according to trainee skill. There are two types of guiding DAS (Fig. 6):

(1) Guiding to the desired behavior, meaning that the control error is reduced by giving some guidance at the operational level, including haptic guidance in steering/pedal operation to enable to follow a desired trajectory.

(2) Teaching the timing needed to perform an action is another guiding option. This involves haptic guidance or using aural signals to guide the driver when steering action is needed in reverse parking.

Shared control can be employed for guiding, in which the strength of the automatic control can be adjusted to change the difficulty. In cases in which the human operator decides to follow the automated system without resistance and/or the operator's intent coincides with that of the system, the strength of the automation control, or level of haptic authority (LoHA) (Abbink et al. 2012) is increased to enhance the supporting effect for the given task. This reduces the task difficulty and the resulting workload of the human operator. On the other hand, the strength of automatic control can increase the task difficulty if the operator intends to resist the automation control, for example, in cases in which the intent of the automated system differs from that of the operator. Under this schema, the consistency in intent between the two agents strongly affects the relationship between the extent of assistance and task difficulty. A method to control cooperative status by changing the DAS's intent to match that of the driver has also been proposed (Nishimura et al. 2015).

**Table 2.** Relationship between training concepts and human machine cooperation

| Training concepts | Types of assistance | Examples | Method for TDA |
|---|---|---|---|
| a) Training in parts = training in one of a set of divided subtasks | Partitioning (1) Fractionization into multiple tasks (2) Segmentation of a task: -Supporting decision making -Replacing in control operation | -Human and DAS perform longitudinal and lateral control, respectively. -Facilitating decision-making; which gap on main lane he/she will merge. -DAS performs lane change by the driver's decision. | Basically, the task to be trained is left when it involves appropriate difficulty. |
| b) Simplifying = lowering a goal of task | Simplifying (1) Giving support when encountering difficulty (2) Simplifying the traffic environment | -DAS works only if it judges that the driver encounters difficulty. -Recommending a vehicle velocity to provide decision-making in a merging position (Suehiro et al. 2018). | The difficulty is defined and the threshold is adjusted based on it. |

| c) Guiding = supporting trainee to reduce errors in task | Guiding (1) Guiding to a desired behavior (2) Teaching the steps necessary to perform a correct action | -Haptic guidance via steering/pedal to enable to follow a given desired trajectory. -Haptic guidance or using sound to tell the driver when steering action is needed in reverse parking. | Strength of automatic control is adjusted. See change of difficulty by timing. |
|---|---|---|---|

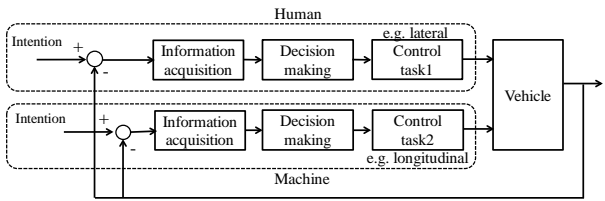

**Fig. 3** Fractionization of multiple tasks

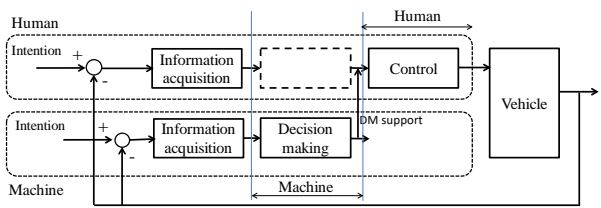

**Fig. 4** Example of segmentation of a task

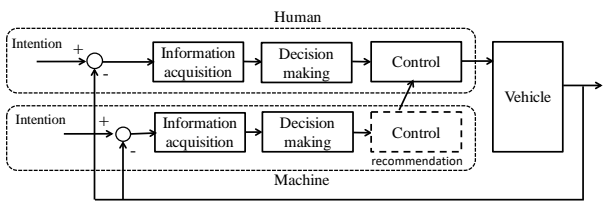

**Fig. 5** Example of traffic environment simplification. If the DAS knows that the driver encounters difficulty in a given scenario and how to decrease this difficulty through control of the ego-vehicle's motion, it provides recommendations for controlling the vehicle motion to simplify the environment (Suehiro et al. 2018).

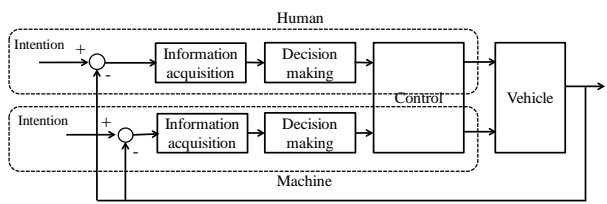

**Fig. 6** Schematic of information processing in shared control. Human and machine/DAS have separate

intentions. In shared control, the control implementation phase is executed by both simultaneously. To execute the operation simultaneously, both the human and DAS engage in information acquisition and decision-making independently. Shared control can perform guidance toward desired behavior and teaching the timing needed to correctly perform actions.

3.5 Interpretation of possible implementation of DASs as training method

3.5.1 Use of shared control

We turn to examples of possible implementation of DASs using haptic shared control to interpret them as a training method.

Reverse parking assistance systems using haptic shared control have been proposed (Hirokawa et al. 2014; Tada et al. 2016), in which the DAS generates the desired path to be followed and the steering wheel provides torque to the driver to follow it. Haptic guidance feeds back the tracking error to the driver in a timely manner and reduces the tracking error. In this manner, HSC operates via guiding training, as discussed in section 3.1.

If the HSC is designed so that the automated system by itself cannot achieve parking, the driver will be encouraged to participate in the control loop actively. As described in section 3.1, excessive feedback during training can increase cognitive load and thereby decrease or eliminate training effectiveness. However, because haptic signaling is a high-speed sensory modality that works directly with the driving operation through the steering, real-time error feedback through HSC can effectively be used without increased cognitive load.

As another example, we examine merging operation assistance in which the driving workload is reduced. The difficulties in merging are thought to arise from the need for parallel execution of decision making, as to which gap the driver will enter while maintaining longitudinal control of the vehicle (Ueda and Wada 2015). As a solution, haptic guidance through the use of a pedal that displays a desired velocity to enter the merging position in the main lane is considered. As the cognitive load is reduced through guidance of longitudinal control, such a system would be expected to increase the longitudinal control skill of a person who is not good at the task. It is also possible that the spare working memory increase obtained through the control assistance would be assigned to enhance decision-making.

As described in the work of F. Flemisch et al.) 2016, shared control can be understood to be the sharp end of cooperation, which means that a connection at the operational level is a consequence of a loose connection between strategic and tactical levels. Thus, interactions at the operational level can be affected by conflict on other levels (Itoh et al. 2016), leading to increases in skill at levels beyond the operational level.

3.5.2 Use of cooperative control

In cooperative control, TDA can be interpreted as part-task training from the viewpoint of training, as the DAS is adjusted by taking over some of the driving. For example, when the DAS assists decision-making in finding a merging position in a situation involving high cognitive load, the driver can concentrate on longitudinal control with a reduced cognitive load on their decision making (Fig. 4). On the other hand, a DAS that recommends driving velocities that decrease difficulty has been proposed (Suehiro et al. 2018) for use in conjunction with methods that evaluate the difficulty of merging position decision-making felt by a driver in a given situation (Ueda and Wada 2015) (Fig. 5). This can be understood to be fractionization-type part-task training, through which skill improvement at the tactical and operational levels is expected to be attainable by simplifying decision-making and longitudinal control, respectively. As described above, structures involving a DAS with skill training and a TDA method with multiple choices are feasible. A systematic method to implement these would be useful, as such decisions tend to be made by trial-and-error based on a task analysis for each given task, which requires significant effort.

4. Examples of simultaneous achievement of workload reduction and skill improvement

A reverse parking assist system using HSC is presented as an example of the simultaneous achievement of assistance and driving skill improvement, which corresponds to point (1)—guiding to a desired behavior of the guiding-type assist system—in section 3.4. See (Tada et al. 2016) for the details of the experiments.

4.1 Overview of the Assistance System

The system assists the driver in reverse parking by applying steering torque to help drive the vehicle into a predetermined parking place after the driver manually stops at the parking starting point (Fig. 7).

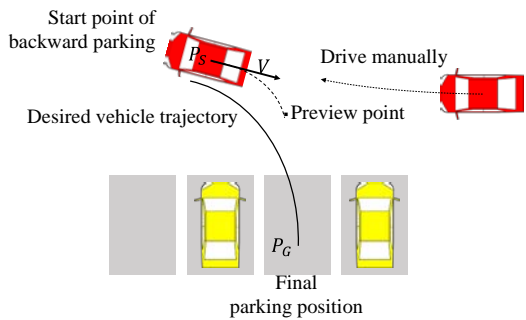

**Fig. 7** Overview of reverse parking assistance by HSC

After the reverse parking starting point is determined, the desired vehicle trajectory is generated and a steering torque is applied to the driver so that the vehicle trajectory error is reduced as follows:

$$\tau_{das}(t) = -C_s |e(t)| \{\theta(t) - \theta_d(t)\} \qquad (5)$$

where $e(t)$ is the error between the current position of the vehicle and its desired trajectory, $\theta(t)$ is the steering angle, and $\theta_d(t)$, which is called the desired steering wheel angle, is the vehicle trajectory error calculated by a second-order preview driver model to decrease the vehicle trajectory error. The desired vehicle trajectory is designed using a third-order Bezier curve with the shortest path length that satisfies the minimum rotational radius of the given vehicle.

4.2 Experimental Method

4.2.1 Design

The effect of the gain settings of the proposed assist system in (5) on changes in driver performance during and after use of the system was investigated. The independent variable of the experiments was the Cs gain, and there were three levels of gain condition: A, with Cs = 0; B, with Cs = 0.5; and C, with Cs = 1.0. Condition A corresponds to no assistance system. Eighteen subjects (fifteen males and three females) aged 19 to 23 years and having driver's licenses participated in the experiments. Each subject drives a vehicle less than once a month. As the gain condition was a between-subject factor, each of the subjects was assigned to only one gain condition; thus, six subjects were assigned to each condition.

4.2.2 Apparatus

Experiments were conducted using a stationary driving simulator (DS), which has four LCD displays at the front and sides, and one at the rear. Drivers were able to acquire information about the backward side with the help of side mirrors displayed as graphics windows on the monitors as well as the graphics windows displayed on the rear monitor through a real rear-view mirror. A 250-W brushless DC motor (Maxon Precision Motors Inc.) was attached to the steering shaft to generate torque around the axis. Computer graphics were generated using Unity 3D (Unity Technologies).

4.2.3 Experimental Procedure

The subjects participated in 26 trials, which are listed in Table 3, after several practice trials. In the a) before-assist phase, subjects performed reverse parking without assistance to determine the driver's initial skill in performing the maneuver. In the before-assist phase, the participants drove forward from an initial position and had to determine where to change their driving direction from forward to backward; this is called the self-selected starting point of backward driving. In the b) during-assist phase, subjects performed the parking operation with assistance control under each given gain condition. In the during-assist phase, participants started backward driving from a fixed point and did not drive in the forward direction. In the c) after-assist with fixed point phase, the subjects again performed assistance-free parking from the same backward driving starting point as in the during-assist phase to evaluate the extent of the increase in their driving skills. As in phase b), the driving trials started backward from a fixed point without forward driving. In the d) after-assist with self-selected point phase, parking was again performed without assistance from a self-selected starting point to determine the extent to which the driver's skills increased. Please note that the initial position of the ego-vehicle and location of other vehicles were fixed throughout the experiments.

Under conditions B and C, subjects were asked to execute reverse driving by following the steering torque generated by the assistance system and not to oppose the system in the during-assist and after-assist with-fixed-point phases.

**Table 3.** Driving phases in experiments

| Phase | Trial | Condition | | Start point |
|---|---|---|---|---|
| | | A | B, C | |
| a) Before-assist | 1-10 | No assist | No assist | Self-selected |
| b) During-assist | 11-20 | No assist | Assist | Fixed |
| c) After-assist with fixed point | 21-23 | No assist | No assist | Fixed |
| d) After-assist with self-selected point | 24-26 | No assist | No assist | Self-selected |

4.3 Results

As shown in Fig. 8, which shows examples of the vehicle trajectory of a subject, the after-assist vehicle trajectories were shorter and smoother than the before-assist trajectories. Fig. 9 shows the vehicle trajectory error for each gain condition. The two-way ANOVA of the mean of the RMS trajectory error by gain setting and driving phase condition, as well as the interaction between these, show that the main effects were significant in the driving phase ($F(3, 20) = 21.352$, $p = 0.000$) and interaction ($F(6, 20) = 2.626$, $p = 0.029$), whereas they were only marginally significant for the gain-setting condition ($F(2, 15) = 2.081$, $p = 0.081$).

The one-way ANOVA of the RMS trajectory error by driving phase for each gain condition shows that the simple main effects were significant in conditions B ($F(3, 20) = 7.425$, $p = 0.002$) and C ($F(3, 20) = 8.424$, $p = 0.001$), but not in condition A ($p = 0.057$).

Post hoc test using the Bonferroni method revealed that the error was significantly smaller during-assist and after-assist with fixed point than before-assist ($p = 0.003$, $p = 0.005$) in condition B. In condition C, the test showed that it was significantly smaller during-assist, after-assist with fixed point, and after-assist with self-selected point than before-assist ($p = 0.001, 0.007, 0.048$). For condition A, the post hoc test showed that error after-assist with fixed point was marginally smaller than after-assist with self-selected point ($p = 0.070$). These results strongly suggest that reverse parking performance significantly improved during assist and the skill improved after assist.

Fig. 10 shows the relationship between the decrease in vehicle trajectory error during assist and that after assist based on the before-assist phase. A positive correlation was found between them ($r = 0.730$, $p = 0.007$).

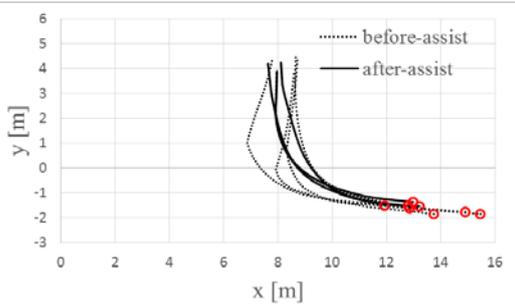

**Fig. 8** Vehicle trajectory

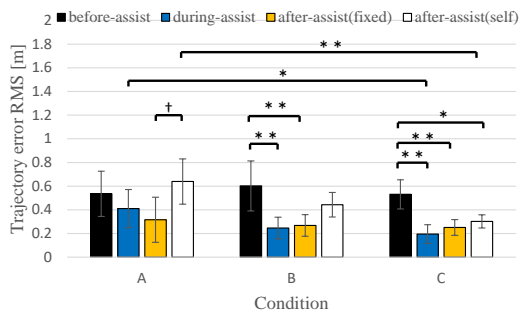

**Fig. 9** Vehicle trajectory error

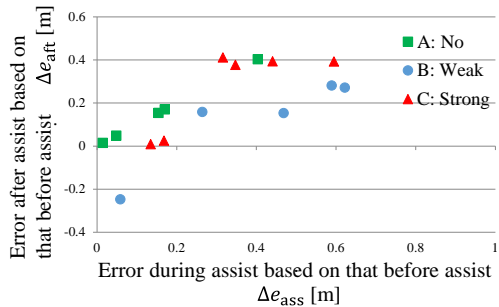

**Fig. 10** Relationship between decrease in error during and after assist

4.4 Discussion

The results regarding vehicle trajectory error shown above and those indicating that the driver's torque was significantly decreased during and after the assist (Tada et al. 2016) strongly suggest that it is possible to simultaneously achieve performance increase and workload reduction during the assist, along with skill increase after the assist. In addition, larger gain settings resulted in higher efficacy both during and after assist. Furthermore, the positive correlation between the decrease in error during and after assist suggests that the larger performance increase during assist facilitates improvement after the assist. The results suggest the importance of adjusting the extent of the assist and the task difficulty for individual skill levels. It should be noted that surrounding vehicle environment and initial position of the ego-vehicle were not changed; thus, the generalization of the training effect to different settings should be investigated. In the experiments, subjects were asked to follow haptic guidance, with the result that intent consistency was apparently maintained. In practice, intent consistency can be violated; in such cases, the cooperative status should be evaluated. Note that, in their pioneering work, Hirokawa et al. (2014) also demonstrated skill improvement following the use of an assistance system; however, they did not report workload reduction. Details of their experimental results and further discussion including the potential factors related to skill enhancement are given in the work of Tada et al. (2016).

5. Conclusion

In this paper, we discussed a methodology for the simultaneous achievement of driver support and skill improvement in shared and cooperative control of DASs. In particular, the importance of task difficulty adjustment was emphasized and the implications of this in shared and cooperative control were discussed on the basis of the relationship between the important factors in training and shared and cooperative controls. Furthermore, research on a reverse parking assistance system employing HSC was introduced as an example of the simultaneous achievement of assist and skill improvement via shared control. The results showed that the drivers' parking performance was significantly improved and that the steering torque was lowered during use of the system; the effect remained even after the assist, in a situation in which there was no support from the system. This indicates that performance improvement with reduced workload and skill improvement were simultaneously achieved through appropriate gain setting.

Concepts of mode (Rieger and Greenstein 1982), form (K. Schmidt 1991), and structure of the cooperation (Millot and Mandiau, 1995) have been introduced to investigate human-machine cooperation, and the synthesis method has been proposed (Pacaux-Lemoine and Flemisch 2016). Discussing the relationship between these concepts and types of assistance introduced in the present study from the viewpoint of the training concepts, is an important future work. In addition, the methodology proposed here will be applied to many types of cooperative DASs to reflect the many varieties of implementation of cooperative control, e.g., the Horse-metaphor (F. O. Flemisch et al. 2003). It is expected that the ideas for discussing the shared and cooperative control of DAS from the viewpoint of training efficacy, workload reduction, and skill development introduced here can be applied to many other contexts. Currently, setting the task difficulty relies strongly on the analysis results of a given task and tends to be performed through trial-and-error. A systematic approach toward developing a procedure for task difficulty adjustment is therefore an important future research topic.

Acknowledgment

This work was partially supported by a JSPS KAKENHI Grant Numbers 26242029 and 17H00842. The authors would also like to thank the anonymous reviewers for their valuable comments and suggestions for improving the quality of the paper.

Sheridan TB (1992) Telerobotics, automation, and human supervisory control. MIT Press.

Suehiro Y, Wada T., Sonoda K (2018) Assistance Method for Merging by Increasing Clarity of Decision Making. IEEE Trans Intell Veh (accepted).

Sweller J (1994) Cognitive load theory, learning difficulty, and instructional design. Learn Instr 4(4):295–312. https://doi.org/10.1016/0959-4752(94)90003-5

Tada S, Sonoda K, Wada T. (2016) Simultaneous Achievement of Workload Reduction and Skill Enhancement in Backward Parking by Haptic Guidance. IEEE Trans Intell Veh 1(4):292–301. https://doi.org/10.1109/TIV.2017.2686088

Ueda S, Wada T (2015) Modeling Driver's Skill of Merging Operation toward Its Assistance System. In Proceedings of the 3rd International Symposium on Future Active Safety Technology Towards zero traffic accidents (pp. 329–334).

Wada T, Sonoda K, Okasaka T, Saito T (2016) Authority transfer method from automated to manual driving via haptic shared control. In 2016 IEEE Int Conf Syst Man Cybern (SMC) (pp. 2659–2664). Budapest: IEEE. https://doi.org/10.1109/SMC.2016.7844641

Wada T, Sonoda K, Tada S (2016) Simultaneous Achievement of Supporting Human Drivers and Improving Driving Skills by Shared and Cooperative Control Achievement of Supporting Drivers and Improving. In Proceedings of 13th IFAC/IFIP/IFORS/IEA Symposium on Analysis, Design, and Evaluation of Human-Machine Systems.

Wada T, Takeuchi T. (2008) A Training System for EMG Prosthetic Hand in Virtual Environment. In Proceedings of Human Factors and Ergonomics Society 52nd Annual Meeting 52:2112–2116. https://doi.org/10.1177/154193120805202706

Wada T, Yoshimura K, Doi S, Youhata H, Tomiyama K (2011) Proposal of an eco-driving assist system adaptive to driver's skill. In 2011 14th International IEEE Conference on Intelligent Transportation Systems (ITSC) (pp. 1880–1885). https://doi.org/10.1109/ITSC.2011.6083034

Wei Y, Bajaj P, Scheidt R, Patton J, Scheldt R, Patton J (2005) Visual Error Augmentation for Enhancing Motor Learning and Rehabilitative Relearning. In 9th International Conference on Rehabilitation Robotics 2005:505–510). https://doi.org/10.1109/ICORR.2005.1501152

Wickens CD, Hollands JG (2000) Engineering psychology and human performance, 3rd Ed., Prentice Hall.

Wickens CD, Lee JD, Liu Y, Gordon Becker SE (2003) An Introduction to Human Factors Engineering, 2nd ed., Pearson Prentice Hall, Upper Saddle River, New Jersey.

Wightman DC, Lintern G (1985) Part-Task Training for Tracking and Manual Control. Hum Factors: J
20